\documentstyle[12pt,cite,epsfig,axodraw,a4]{article}

\newcommand{\la}{\langle}
\newcommand{\ra}{\rangle}

\def\E{\mbox{e}^+\mbox{e}^-}
\newcommand{\beq}{\begin{equation}}
\newcommand{\eeq}{\end{equation}}

\begin{document}

\hfill Nijmegen preprint

\hfill HEN-394

\hfill June 1996

\vspace{1.0cm}

\begin{center}

{\Large\bf
Local Multiplicity Fluctuations in Z decay\footnote{
Presented at 7th International Workshop on
Multiparticle Production ``Correlations and Fluctuations'',
Nijmegen, The Netherlands, June 30 - July 6, 1996}}

\vspace{1.5cm}

{\large S.V.Chekanov\footnote{On leave from
Institute of Physics,  AS of Belarus,
Skaryna av.70, Minsk 220072, Belarus}}

\medskip

{\it High Energy Physics Institute Nijmegen
(HEFIN), University of Nijmegen/NIKHEF, \\
NL-6525 ED Nijmegen, The Netherlands}

\vspace{1.2cm}

for

\vspace{1.2cm}

{\bf\large L3 Collaboration}

\vspace{1.3cm}

\end{center}

\begin{abstract}
Local  multiplicity  fluctuations of
hadrons produced in the decay of $Z^0$
were studied on the basis of L3 data.
In addition to the normalized-factorial-moment method,
the fluctuations were studied for the first time
by the use of bunching parameters.
A strong multifractal structure was observed
inside jets.
JETSET 7.4 PS describes the fluctuations
in the azimuthal angle defined with
respect to the beam axis reasonably well.
For the fluctuations in rapidity, defined with
respect to the thrust axis and in the
four-momentum-difference variable,
JETSET 7.4 PS overestimates the fluctuations.
\end{abstract}

\newpage

\section{Introduction}
\label{sec:int}

The quest for
local multiplicity 
fluctuations is the quest for short-range 
correlations
in a  multiparticle system in which particles
have a tendency to form
so-called ``spikes'' according to underlying 
stages of the multiparticle process.
In high-energy physics, spikes are seen as
dynamical peaks in the
phase-space distribution of individual events.
The dynamical occurrence of spikes leads to 
the intermittency phenomenon \cite{1,4,5},
defined as a power-law behavior
of the normalized factorial moments (NFMs) 
\beq
F_q(\delta )=\frac{\la n^{[q]}\ra}
{\la n\ra^q}\propto\delta^{-\phi_q},
\qquad n^{[q]}=n(n-1)\ldots (n-q+1),
\label{1}
\eeq 
where $n$ is the  number of particles 
in a restricted phase-space interval
of size $\delta$, $\la\ldots\ra$ is 
the average over all events in the sample, 
and $\phi_q>0$ is the intermittency index.

Of course, this method of local-fluctuation analysis is not unique.
The fluctuations can be investigated by any quantity characterizing
the multiplicity distribution in $\delta$, if we know {\em a priori} 
its behavior
in the case of statistical fluctuations, i.e., 
when the occurrence of spikes
is caused by purely statistical reasons. 
In the simplest approach,
one can restrict oneself to those quantities 
which have $\delta$-independent behavior
for purely statistical phase-space fluctuations, 
as is the case for NFMs.   

From a theoretical point of view,  the NFMs have the
additional advantage that they 
filter out Poissonian noise \cite{1}. This
property is of vital importance for the
comparison of  
theoretical models involving
an infinite number of particles 
in an event to the experimental data.

Bunching parameters (BPs) 
have  similar  properties for the local-fluctuation study  
\cite{6,7,8,9}.
The definition of the BPs is
\begin{equation}
\eta_q(\delta )=\frac{q}{q-1}\frac{P_q(\delta )P_{q-2}(\delta )}
{P_{q-1}^2(\delta )},
\label{2}
\end{equation}
where $P_n(\delta )$ is the probability of having $n$ particles inside
a restricted phase-space interval of size $\delta$.
The main advantage of these quantities over the NFMs is that  they 
are more sensitive to the structure of local
fluctuations \cite{9}. Another property is that
for multifractal local fluctuations,  $\eta_q(\delta )$ is a
$\delta$-dependent function  for all $q>2$, 
while for monofractal behavior, one has
$\eta_q(\delta ) =const$ for $q>2$ \cite{6}. 
This property simplifies 
the multifractal analysis significantly: 
any observation of the power-law
$\delta$-dependence of 
$\eta_q(\delta )$ for $q=3, 4, \ldots$ means that the 
anomalous fractal dimension $d_q=\phi_q/(q-1)$ has a tendency 
to increase with increasing $q$,  so that  the sample 
exhibits a multifractal property.

In addition, from an experimental point of view,
BPs have the following two important properties \cite{9}:
1) They are less severely affected by
the bias from finite statistics
than are the NFMs, since the $q$th-order BP
resolves only the behavior
of the multiplicity distribution near multiplicity $n=q-1$;
2)  For the calculation of the BP of order $q$,
one needs to know only the $q$-particle 
resolution of the detector, 
not any higher order resolution. 

In this paper, we present an experimental 
investigation of 
local fluctuations in the final-state hadron system produced
in $Z^0$ decays at $\sqrt{s}=91.2$ GeV  by using 
both NFMs and BPs. 
The  final-state hadrons have been 
recorded with the L3 detector during
the 1994 LEP running period.
The calculations are based 
on approximately 810k selected hadronic events.    
Charged hadrons 
are selected by  the standard L3 selection procedure, based on
energy deposition  in the electromagnetic
and hadronic calorimeters and
momentum measurement in the
Central Tracking Detector including the 
L3 Silicon Microvertex Detector. 

\section{Experimental definitions}
\label{sec:exp}

In order to  increase the statistics and to reduce
the statistical error in observed local quantities
when analyzing experimental data,
we use the bin-averaged NFMs and BPs \cite{9}:

\medskip

1) {\it Horizontal NFMs}:
\beq
F_q(\delta )=\frac{1}{M}\sum_{m=1}^{M}
\frac{\la n_m^{[q]}\ra}{\la \bar n \ra^q},
\label{18lo}
\eeq
where $n_m$ is the number of particles in bin $m$, 
$\la \bar n \ra =\bar N/M$, $\bar N$ is the average
multiplicity for full phase space,
$M=\Delta /\delta$  is the total number of bins, and
$\Delta$ represents the full phase-space volume.

\medskip
2) {\it Horizontal BPs}: 

\begin{equation}
\eta_q(\delta )=
\frac{q}{q-1} \frac{\bar N_q(\delta ) \bar N_{q-2}(\delta )}
{\bar N_{q-1}^2(\delta )}, \qquad 
\bar N_{q}(\delta )=\frac{1}{M}\sum_{m=1}^{M}N_q(m,\delta ),
\label{3}
\end{equation}
where $N_q(m,\delta )$ is the number of events  having $q$ particles
in  bin $m$, $M=\Delta /\delta$.

Both quantities (\ref{18lo}) and (\ref{3}) are equal to unity for
a purely independent particle production following
a Poissonian multiplicity distribution in restricted bins.

Note that the definitions presented above can be used in practice 
for a flat single-particle density distribution. 
To be able to study  non-flat distributions,
we carry out a transformation from the original
phase-space  variable to that
in which the underlying 
distribution is approximately uniform \cite{10,11}. 

\medskip
3) {\it Generalized integral BPs:}

Recently, a new set of bunching-parameter 
measurements has been proposed that make use of the 
interparticle-distance measure  technique \cite{9}.
To study fluctuations of spikes, we will
consider the generalized integral BPs 
using  the pairwise squared 
four-momentum difference $Q^2_{12}=-(p_1-p_2)^2$.
The definition of the BPs is given by
\begin{equation}
\chi_{q}(Q^2_{12})=
\frac{q}{q-1} \frac{\Pi_q(Q^2_{12}) \Pi_{q-2}(Q^2_{12} )}
{\Pi^2_{q-1}(Q^2_{12})}, 
\label{7we}
\end{equation}
where $\Pi_i (Q^2_{12})$ represents the number
of events having  $i$  spikes of size $Q^2_{12}$, irrespective of
how many particles are inside each spike.
To define the spike size, we  used
the so-called Grassberger-Hentschel-Procaccia (GHP)
counting topology
\cite{z15,z16}, for which a $g$-particle hyper-tube is assigned
a size $\epsilon=Q^2_{12}$ that corresponds to the
maximum of all pairwise distances.
For purely independent particle production with the
spike multiplicity distribution  characterized by a Poissonian
law, the BPs (\ref{7we})
are equal to unity for all $q$.

\section{Analysis}
\label{sec:anal}

Two samples of multihadronic events are 
generated with JETSET 7.4 PS.
The first sample contains all  charged
final-state particles with a lifetime greater than
$10^{-9} \mathrm{s}$ (generator-level sample).
This sample is generated with initial state photon radiation.
The second, detector-level sample,  includes distortions due
to detector effects, limited acceptance, finite resolution
and the event selection.
Both the generator-level  and
detector-level samples have the same 
statistics (810k hadronic events).

A corrected NFM or BP is found by means of the following
correction procedure
\begin{equation}
D_{q}^{\mathrm{cor}}=C_{q} \> D_{q}^{\mathrm{raw}}, \qquad
C_q=\frac{M_q^{\mathrm{gen}}}
{M_q^{\mathrm{det}}}.
\label{da1}
\end{equation}

Here, $M_q^{\mathrm{gen}}$ and $M_q^{\mathrm{det}}$ symbolize
an NFM or BP of order $q$ calculated from the generator-level
and detector-level Monte-Carlo samples, respectively. 
$D_q^{\mathrm{raw}}$ represents an NFM or
BP  calculated directly from the raw data.
The same correction procedure has been used in
\cite{12,13}.

\subsection{In the detector frame}
\label{sec:det}

\begin{figure}[htb]
\begin{center}~
\begin{picture}(435,230)
\put(-25,20){
         {\epsfig{file=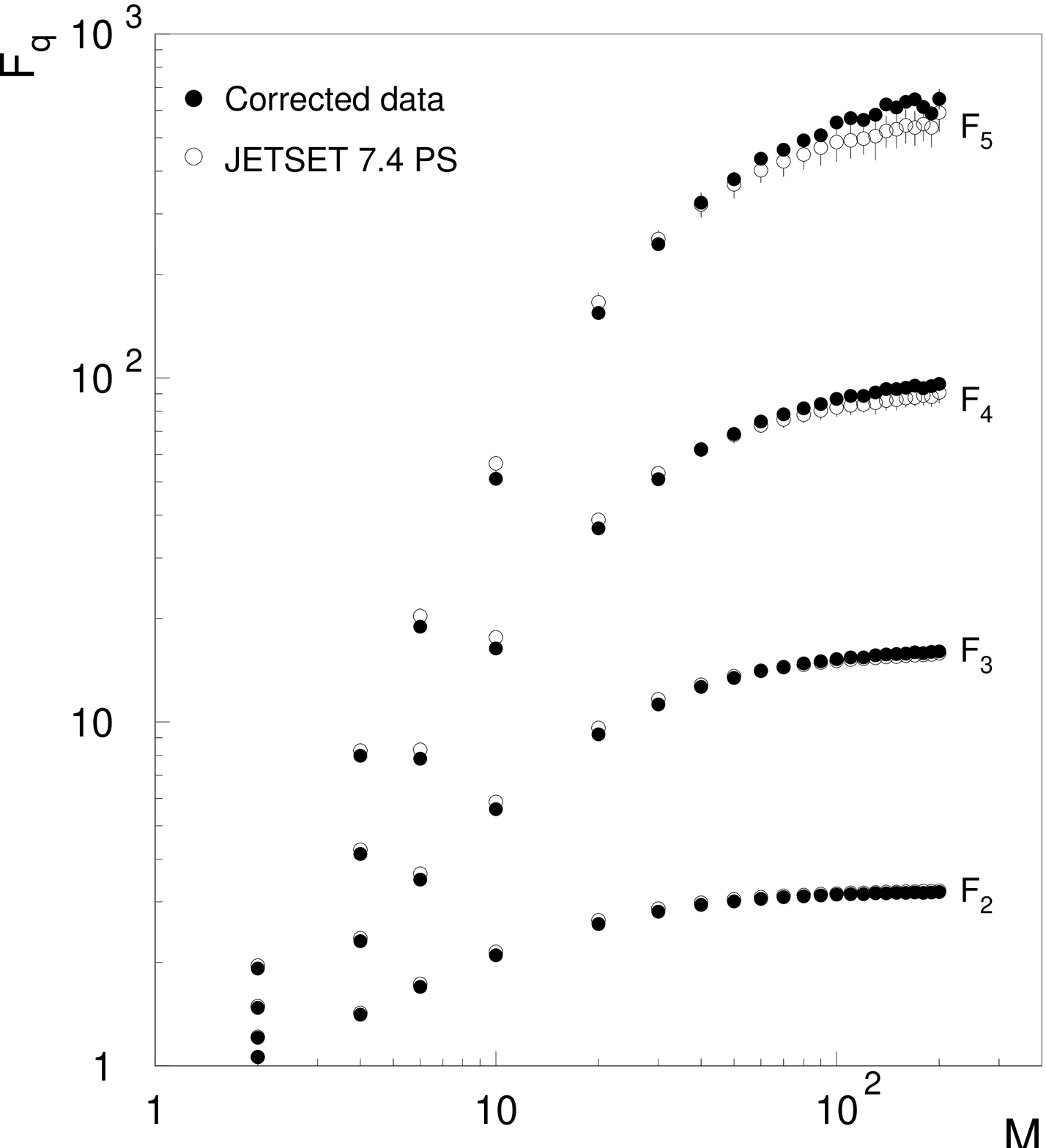,width=210pt,height=310pt}}
         }
\put(200,20){
         {\epsfig{file=bp_az_b.eps,width=220pt,height=260pt}}
         }
\end{picture}
\end{center}

\begin{minipage}[t]{6.5cm}
\caption[fig1]{
NFMs  as a function of
the number $M$
of bins in
azimuthal angle  defined
with respect to the beam axis.}
\label{fig1}
\end{minipage}
\hspace{1.4cm}
\begin{minipage}[t]{6.5cm}
\caption[fig2]{
BPs as a function of
the number $M$
of bins in
azimuthal angle  defined
with respect to the beam axis.}
\label{fig2}
\end{minipage}

\end{figure}

To update the results already 
presented in \cite{14}, we carried out   
measurements of  horizontal NFMs (\ref{18lo}) 
as a function 
of the number $M=2\pi /\delta\varphi$
of partitions of the full angular interval $2\pi$, 
where $\delta\varphi$ denotes the bin size in azimuthal
angle $\varphi$ defined
with respect to the beam axis. 
Since the event averaged distribution in
$\varphi$ is uniform, the Ochs-Bia\l as-Gazdzicki 
transformation \cite{10,11}
was not performed here. 
In Fig.~\ref{fig1}, 
the corrected data
are shown by full symbols and 
the generator-level of JETSET 7.4 PS tuned by the L3 Collaboration 
by open symbols.
Here and below, the smallest bin size is estimated from the Monte-Carlo
study of the charged-track resolution of the L3 detector in the
particular variable.

The statistical errors on the data shown in Fig.~\ref{fig1}
are derived from the covariance matrix
of the horizontally averaged factorial moments.
They include
the statistical error on
the correction factor $C_q$ in (\ref{da1}).
To combine the statistical error on the correction
factor, we assume that the statistical 
errors for the generator-level
and detector-level Monte Carlo's are independent. This (strong) assumption
leads to an upper limit of the error derived for $C_q(\delta )$.

The error bars on the Monte-Carlo predictions
include both statistical and systematical
errors. The systematical errors have been estimated
by varying, by one standard 
deviation, the LUND fragmentation parameter PARJ(42), 
the width of the Gaussian $p_x$ and $p_y$ hadronic transverse
momentum distribution (PARJ(21)),
and the $\Lambda$ value used for $\alpha_s$ in
parton showers (PARJ(81))\footnote{The 
value of these parameters
have been  tuned by the L3 Collaboration to reproduce the
single-particle spectra and global-shape distributions.}.
For the given statistics, the errors on Monte Carlo are
dominated by the systematical errors, so that
the open symbols represent the
values of NFMs with the L3 default and
the error bars indicate the maximum and 
minimum values  obtained after
the parameter variations.

Fig.~\ref{fig1} shows that the Monte-Carlo predictions 
slightly oscillate  around the corrected
data, but reasonably reproduce the experimental results.

The same hadronic sample is used to calculate
the horizontal type of the BPs (\ref{3}).
The behavior of $\ln \eta_q(M)$ as a function of
$\ln M$ is presented in Fig.~\ref{fig2}.
Being more sensitive to the structure of  fluctuations,
the BPs show that 
JETSET 7.4 PS slightly overestimates the increase  of the
second-order BP and oscillates around the third-order BP calculated
from the data. The Monte-Carlo 
predictions reproduce the
higher-order BPs reasonably well. 
The observed decrease of the high-order BPs
with increasing $M$ reflects a particle
antibunching of the order $q>2$  due to jet structure, i.e., 
particles are bunched together inside each jet, but there are
phase-space intervals between the bunches due to energy-momentum
conservation.

\subsection{In the event frame}
\label{sec:eve}

\begin{figure}[htb]
\begin{center}~
\begin{picture}(435,230)
\put(-25,20){
         {\epsfig{file=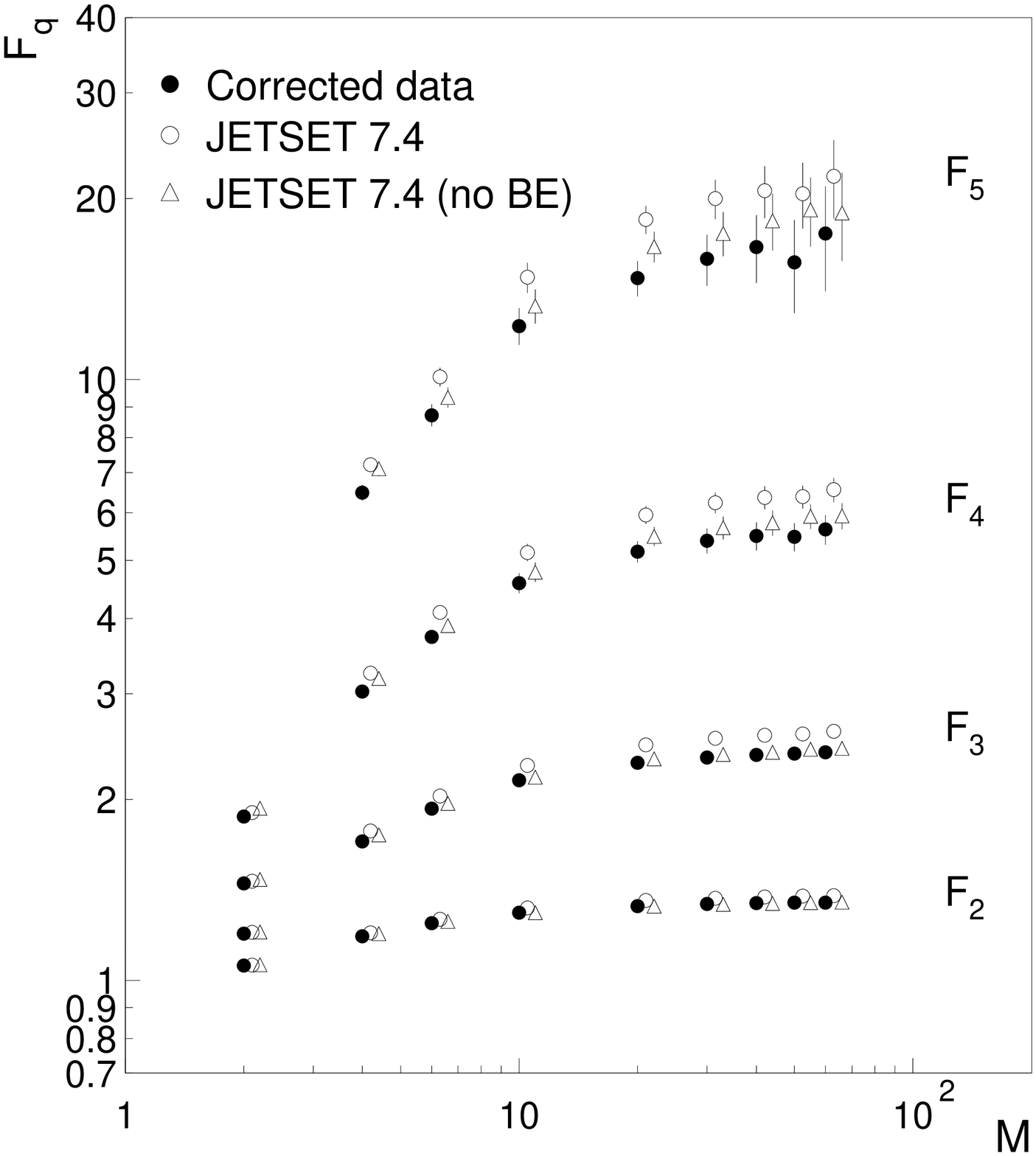,width=210pt,height=310pt}}
         }
\put(200,20){
         {\epsfig{file=bp_all.eps,width=220pt,height=260pt}}
         }
\end{picture}
\end{center}

\begin{minipage}[t]{6.5cm}
\caption[fig3]{
NFMs  as a function of
the number $M$
of bins in rapidity 
$y$ defined with respect to the thrust axis.}
\label{fig3}
\end{minipage}
\hspace{1.4cm}
\begin{minipage}[t]{6.5cm}
\caption[fig4]{
BPs as a function of
the number $M$
of bins in rapidity 
$y$ defined with respect to the thrust axis.}
\label{fig4}
\end{minipage}

\end{figure}

With the above observation in mind, it is obvious 
that the NFMs and BPs 
calculated so far are strongly
influenced by the jet structure of events.
In order to study genuine fluctuations inside jets,
we used the rapidity $y$ defined
with respect to the thrust axis.
The NFMs as a function of the number of bins in $y$ 
are shown in Fig.~\ref{fig3}. The predictions
of the JETSET 7.4 PS tuned by the L3 Collaboration are 
presented by open circles.
The behavior of the NFMs shows the same trend as that in the
azimuthal angle $\varphi$ defined 
with respect to the beam axis. The
signal observed, however, is much smaller for the present calculations.
As we see, JETSET 7.4 PS 
overestimates the intermittency effect for large $M$.
This discrepancy is  enhanced with rising
moment order.

\vspace{1.2cm}

\medskip

Fig.~\ref{fig4} shows the results for the horizontally normalized
BPs (\ref{3}) in the rapidity $y$ defined 
with respect to the thrust axis.
In contrast to Fig.~\ref{fig2}, all high-order BPs show a
power-law  increase with increasing $M$,
indicating that the fluctuations  in this variable
are of  multifractal type.
The  multifractality observed, therefore, 
appears to be a consequence of the
cascade nature of parton branching, hadronization, resonance decays
and Bose-Einstein correlations.
Note that the conclusion on the multifractal
type of the fluctuations becomes possible without the
necessity for the calculation of the intermittency indices.
In contrast, to reveal multifractality with the help of the NFM-method
one needs to carry out fits of the NFMs by a power law.

A disagreement with the Monte-Carlo  predictions is observed for
$q=2,3$, while higher-order BPs are described well by the model.

As mentioned in \cite{15}, the influence of 
Bose-Einstein (BE) correlations
on quantities measured  depends strongly on the type of quantity and
variable used. Obviously,  the  BE correlations are a 
typical candidate for  the cause of local
fluctuations in $3$-momentum phase
space, which  should lead to a rise of  
fluctuations in one-dimensional
phase space as well.
To demonstrate this effect, Figs.~\ref{fig3} and \ref{fig4} also
show a comparison 
of the JETSET 7.4 PS model without  
BE interference (open triangle symbols) to the data.
Indeed, the model expectations without  the BE effect
in Fig.~\ref{fig3} have a smaller rise of 
the NFMs than do those that include  the BE effect.

It is quite remarkable  
how well the influence of BE correlations on 
the local fluctuations in JETSET
can be seen  in
the second-order BP (see Fig.~\ref{fig4}).
This influence is due to the fact that the BE effect is implemented in
JETSET on the level of  two-particle correlations, which
are strictly related to the second-order BP (NFM).

\subsection{In the four-momentum difference}
\label{sec:fo}

\begin{figure}[htb]
\begin{center}~
\begin{picture}(435,230)
\put(-25,20){
         {\epsfig{file=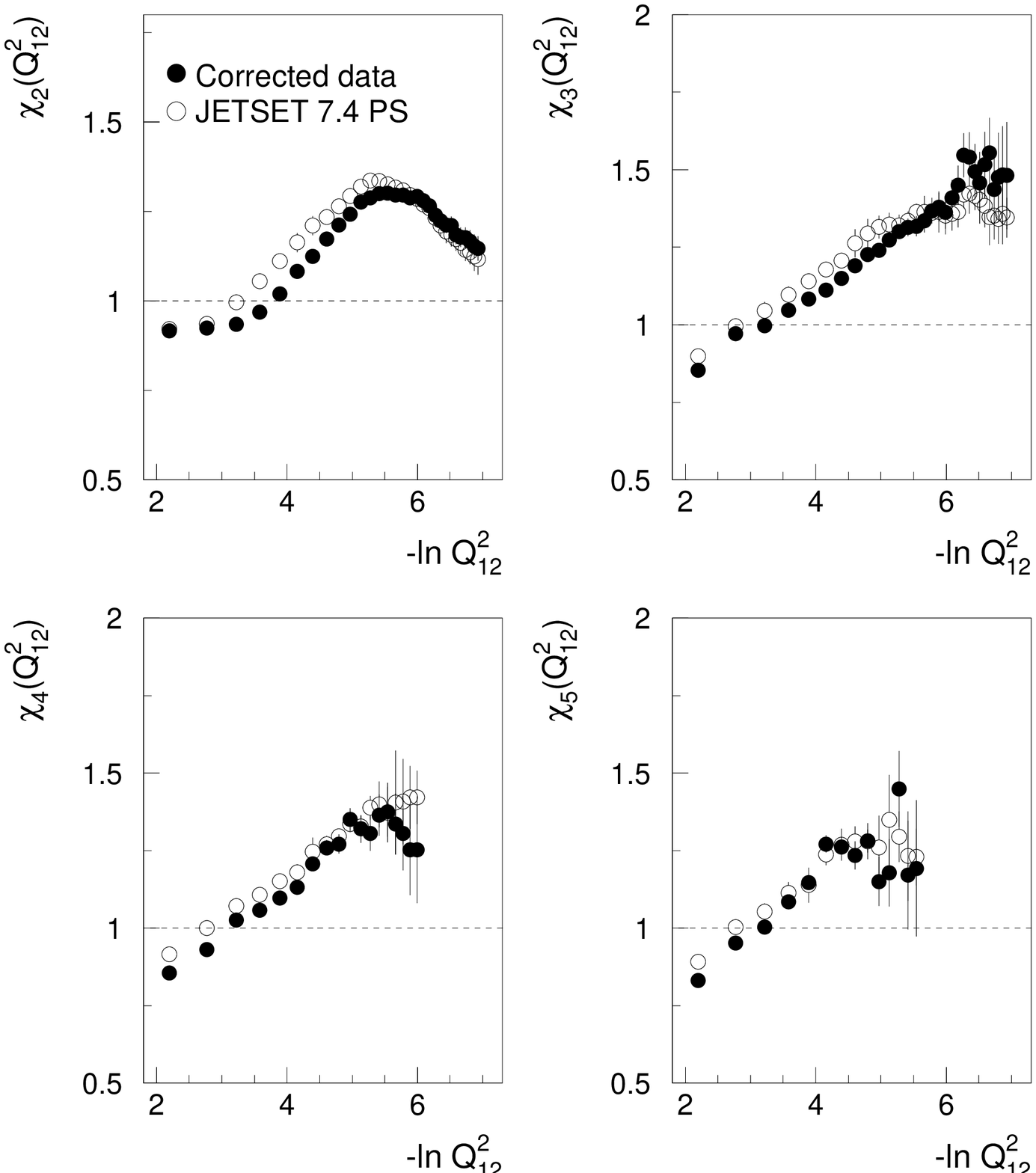,width=210pt,height=260pt}}
         }
\put(200,20){
         {\epsfig{file=int2.eps,width=220pt,height=320pt}}
         }

\end{picture}
\end{center}

\begin{minipage}[t]{6.5cm}
\caption[fig5]{
Generalized integral BPs 
as a function of
the squared four-momentum difference $Q^2_{12}$
between two charged particles.}
\label{fig5}
\end{minipage}
\hspace{1.4cm}
\begin{minipage}[t]{6.5cm}
\caption[fig6]{
Generalized 
second-order BP 
as a function of
the squared four-momentum difference $Q^2_{12}$
between two charged particles.}
\label{fig6}
\end{minipage}

\end{figure}

Fig.~\ref{fig5} shows the behavior of the 
generalized BPs (\ref{7we}) as a 
function of $Q^2_{12}$.
The dashed lines represent the behavior of these BPs
in the Poissonian case. 
To contrast, all BPs rise with decreasing $Q^2_{12}$,
which corresponds to a 
strong bunching effect of all orders, leading
to multifractal fluctuations.
The saturation and
downward bending of
the second-order BP at small $Q^2_{12}$ are
caused by the influence of
resonances at intermediate $Q^2_{12}$ (see below).

\medskip

To investigate this disagreement  in more detail,
we present in Fig.~\ref{fig6}  
the behavior of the second-order
BP as a function of $Q^2_{12}$ for 
multiparticle  hyper-tubes (spikes) made of
like-charged and  that of  unlike-charged particles.
A large difference is observed between these two samples due to
different particle dynamics.
For like-charged particle combinations 
(i.e., for spikes with a maximum charge),
a strong bunching
effect ($\chi_2(Q^2_{12})>1$) is seen at small $Q^2_{12}$.
However, the bunching is much smaller
and even disappears at small $Q^2_{12}$ 
for unlike-charged particle combinations.
This effect can be explained by
resonance decays at intermediate $Q^2_{12}$,
when decay products of short-lived resonances
tend to be separated in the phase space. 
The antibunching effect ($\chi_2(Q^2_{12})<1$) for large $Q^2_{12}$ is
caused by the energy conservation constraint \cite{9}.

The resonance effect is much weaker for like-charged combinations.
In addition, the BE correlations strongly affect the
like-charged particle combinations. Note, however, that
JETSET 7.4 PS leads to a strong rise 
of $\chi_q(Q^2_{12})$ for like-charged particle
combinations, even without the modeling of the 
BE interference. 

JETSET 7.4 PS overestimates the data
for unlike-charged combinations 
at intermediate values of $Q^2_{12}$ and 
underestimates
the data for like-charged combinations at small $Q^2_{12}$.
The latter disagreement
can be reduced, in part, by varying the BE parameters 
in JETSET 7.4 PS\footnote{Note that the study of the BE 
effect for charged particles has not been performed 
by the L3 Collaboration so far.}.
However, we have verified that the  
disagreement for  unlike-charged combinations cannot  be 
reduced by only varying the BE parameters.
 
The most probable shortcomings leading to the discrepancies found
are the simulation of hadronization\footnote{
For $2.5<-\ln Q^2_{12}<5.0$,
our calculations show a large sensitivity of the results
obtained to LUND fragmentation, since
large systematic errors for this domain of $Q_{12}^2$
come mainly from varying the LUND  fragmentation parameter
PARJ(42) and PARJ(21) by one standard deviation.}
and the BE effect. 
As an example, the residual distortion of the decay 
products of short-lived
resonances by BE correlations 
not yet implemented in the
JETSET 7.4 PS model may be a good candidate as an explanation for 
such a discrepancy.
The importance of the latter effect was realized  recently,
when a significant mass shift of $\rho^0$ was observed by
OPAL and DELPHI \cite{17,18}.

The production rate of $f_0(975)$ and $f_2(1270)$ 
measured by DELPHI \cite{18} is 
another challenge for the JETSET model.
In this respect, it is not improbable that a much larger 
fraction of the observed final-state
hadrons results from resonance decays than is usually assumed.
In  this case, the negative correlations should be larger, and
a better agreement  with the data for the 
intermediate values of  $Q^2_{12}$ 
would be achieved for unlike-charged particles.  
Indeed, we have found that a realistic small variation of   
the production of resonances
($\rho$, $\omega$, $\eta$, $\eta '$) 
responsible for the  unlike-charged particle fluctuations
in the JETSET 7.4 PS can lead to a better agreement.   
This is not likely to improve the
discrepancy fully, however, since the 
JETSET 7.4 PS tuned by the L3 Collaboration shows a reasonable
agreement with the production rates of the main resonances 
\cite{20} and the
variation of the parameters should  not  be large.

As an additional verification, the default version of JETSET 7.4 PS
has been compared to the data.
The same kind of the disagreement 
is found (not shown).

Of course, the disagreement for the unlike-charged
particle combinations in $Q^2_{12}$  (and, hence, for
the all-charged combinations shown in Fig.~\ref{fig5})  can 
also lead to the disagreement
between the JETSET 7.4 PS and the data
in the case of the one-dimensional
variables $\varphi$ and $y$ presented in 
subsections ~\ref{sec:det} and  \ref{sec:eve}.

\section{Discussion}

For the first time, local multiplicity 
fluctuations have been 
studied by means of  bunching parameters. 
Using this method, fluctuations in rapidity defined with respect to
the thrust axis  and  in  
the four-momentum difference $Q_{12}^2$  
are found to exhibit  a strong   
multifractal behavior. The multiplicity distributions
in these variables, therefore,
cannot be described by conventional distributions
(Poisson, geometric,
logarithmic, positive-binomial,
negative-binomial), which have
$\delta$-independent high-order BPs \cite{6}.
Recently,  more general multiplicity distributions with  
power-law high-order BPs have been considered \cite{7}. 
Such types of distributions, therefore,
appear to be more relevant to the situation observed. However,
a phenomenological description of these distributions 
can, to only a slight extent,
provide a physical explanation of the nature
of  multifractal behavior.

For an $\E$ interaction, one can be confident
that, at least on the parton level of this reaction,
perturbative QCD calculations  can give a 
hint for understanding the problem.
Analytical calculations based on the double-log approximation 
of perturbative QCD show that the multiplicity distribution
of partons in ever smaller opening angles is
inherently multifractal \cite{7in, 8in, 9in}.
Of course, the choice of variable can affect the observed signal,
and the final conclusion about agreement
between QCD and the data 
can only be derived  after the calculation
of local quantities in angular variables
that are defined with respect to the thrust (or sphericity) axis. 

In part, the disagreement found between JETSET and the data  is due to
problems in  tuning the JETSET 7.4 PS 
model by the L3 Collaboration,
since the tuning of the model has been performed 
by means of global observables only.
However, the problem of the discrepancy found
can  be more complicated, and
an  additional study of JETSET itself is necessary: 
it has been shown that JETSET 7.4 PS overestimates
in the $Q^2_{12}$ variable
fluctuations of spikes made of unlike-charged 
particles. 
 
Thus, it appears that some important point 
in the simulation of hadronization
and BE interference is
missing in the present version of JETSET  and further 
modifications of the model are needed. 
A similar conclusion has been derived in \cite{19fi}, where
it was shown that  JETSET fails to reproduce 
the multiplicity dependence of the intermittency index.

\bigskip
Acknowledgments
\medskip

This work is part of the research program of the ``Stichting voor
Fundamenteel Onderzoek der Materie (FOM)'', which
is financially supported by the ``Nederlandse Organisatie voor
Wetenschappelijk Onderzoek (NWO)''.
I am greatly indebted to 
the CERN accelerator divisions for the excellent 
performance of the LEP machine.
I acknowledge the effort of all engineers and 
technicians who have participated 
in the construction and maintenance of the L3
experiment. 
I thank the experimental department of
HEFIN for the hospitality.
I am grateful to
W.Metzger, V.I.Kuvshinov, A.Buytenhuijs
for helpful discussions.
Special thanks to W.Kittel for numerous and fruitful
discussions on the subject and for a
careful reading of the manuscript.

\end{document}